\documentclass[10pt, conference]{IEEEtran}

\usepackage{caption}
\usepackage{cite}
\usepackage{url}
\usepackage{amssymb}
\usepackage{blindtext}
\usepackage{comment}

\usepackage{subcaption}
\usepackage{xcolor,soul,framed}
\usepackage{amsmath}
\usepackage{graphicx}
\usepackage{multirow}
\IEEEoverridecommandlockouts
\def\BibTeX{{\rm B\kern-.05em{\sc i\kern-.025em b}\kern-.08em
    T\kern-.1667em\lower.7ex\hbox{E}\kern-.125emX}}

\ifCLASSINFOpdf
\else
\fi

\begin{document}
\title{RAILS: Risk-Aware Iterated Local Search for \\Joint SLA Decomposition and Service Provider Management in Multi-Domain Networks}

%University of Amsterdam, LAB42, Science Park 900, 1098 XH Amsterdam, The Netherlands
\author{\IEEEauthorblockN{Cyril Shih-Huan Hsu}
\IEEEauthorblockA{Informatics Institute\\
University of Amsterdam\\
Amsterdam, The Netherlands\\
s.h.hsu@uva.nl}
\and
\IEEEauthorblockN{Chrysa Papagianni}
\IEEEauthorblockA{Informatics Institute\\
University of Amsterdam\\
Amsterdam, The Netherlands\\
c.papagianni@uva.nl}
\and
\IEEEauthorblockN{Paola Grosso}
\IEEEauthorblockA{Informatics Institute\\
University of Amsterdam\\
Amsterdam, The Netherlands\\
p.grosso@uva.nl}
}


%Telephone: (800) 555--1212, Fax: (888) 555--1212}


\maketitle

\begin{abstract}
The emergence of the fifth generation (5G) technology has transformed mobile networks into multi-service environments, necessitating efficient network slicing to meet diverse Service Level Agreements (SLAs). SLA decomposition across multiple network domains, each potentially managed by different service providers, poses a significant challenge due to limited visibility into real-time underlying domain conditions. This paper introduces Risk-Aware Iterated Local Search (RAILS), a novel risk model-driven meta-heuristic framework designed to jointly address SLA decomposition and service provider selection in multi-domain networks. By integrating online neural network (NN)-based risk modeling with iterated local search principles, RAILS effectively navigates the complex optimization landscape, utilizing historical feedback from domain controllers. We formulate the joint problem as a Mixed-Integer Nonlinear Programming (MINLP) problem and prove its NP-hardness. Extensive simulations demonstrate that RAILS achieves near-optimal performance, offering an efficient, real-time solution for adaptive SLA management in modern multi-domain networks.

\end{abstract}


% \fi
\IEEEpeerreviewmaketitle

\section{Introduction}
The advent of 5G has transformed mobile networks into multi-service environments tailored to diverse industry needs. A key enabler of this shift is network slicing, which creates multiple End-to-End (E2E) logical networks over shared infrastructure, each customized per Service Level Agreements (SLAs). SLAs define expected Quality of Service (QoS) through Service-Level Objectives (SLOs), covering metrics like throughput, latency, reliability, and security.
%Since a network slice often spans across multiple segments of the network, including (radio) access, transport, and core networks, involving multiple operators, meeting SLOs requires decomposing the E2E SLA into partial SLAs suited to each domain's capabilities.
A single network slice may span across multiple segments of the network, including (radio) access, transport, and core networks, and it may involve collaboration between different operators and infrastructure providers. To ensure that the service meets the agreed-upon SLOs across these domains, it is essential to adjust the service parameters accordingly. As a result, the E2E SLA associated with a network slice must be partitioned into specific SLOs for each domain.
This decomposition is crucial for effective resource allocation and remains a core challenge in network slicing.
Several studies have discussed this issue. \cite{ietf-teas-5g-network-slice-application-03} highlights the complexity of mapping E2E requirements to transport networks. \cite{hcltech2023networkslicing} focuses on lifecycle automation, orchestration, and real-time monitoring for SLA compliance. \cite{iovanna2022networkslicing} stresses the role of SLA parameters in E2E QoS and the need for appropriate transport resources. Additionally, \cite{su2019resource} underscores the importance of SLA decomposition for resource allocation, while \cite{10011552} explores AI-assisted SLA decomposition in automating 6G business processes.

In typical network slicing management architectures, a two-level hierarchy is employed~\cite{Vleeschauwer21_SLAdecomposition, SLADNN23, hsu2024online}. This includes an E2E service orchestrator, responsible for overseeing the lifecycle management of network services, and local domain controllers, which manage the instantiation of network slices within their specific domains. The orchestrator determines how the E2E SLA is partitioned into domain-specific SLOs. However, a common constraint is that the orchestrator usually lacks real-time visibility into the state of each domain's infrastructure at the moment of decomposition. Instead, it relies on historical data reflecting the outcomes of previous slice requests.
Several studies~\cite{8417711, 8931583, 10173672} have introduced prediction-based approaches for SLA management, though they do not explicitly tackle the E2E SLA decomposition problem. In~\cite{8417711}, the authors proposed a mapping layer that oversees the network within a service area, managing radio resource allocation to slices to ensure their target service requirements are met. The work in~\cite{8931583} presented an SLA-constrained optimization method leveraging Deep Learning (DL) to estimate resource requirements based on per-slice traffic. Similarly,~\cite{10173672} utilized a context-aware approach, employing graph representations to predict SLA violations in cloud computing environments.
Additionally, heuristic-based SLA decomposition methods have been explored in prior research~\cite{su2019resource}. In~\cite{9165317}, the authors introduced an E2E SLA decomposition system that applies supervised machine learning to partition E2E SLAs into access, transport, and core SLOs.

In our previous work~\cite{Vleeschauwer21_SLAdecomposition, SLADNN23}, we tackled the SLA decomposition problem using neural network (NN)-based risk models in a two-step approach that combined machine learning and optimization. Building on that,~\cite{hsu2024online} introduced an online learning–decomposition framework for dynamic, multi-domain SLA management. However, these studies assumed a preselected service provider per domain, focusing solely on optimizing E2E acceptance probabilities.
In real-world network environments, multiple service providers are often available within each domain, offering varying performance characteristics and capabilities.
For example, in a 5G network slice for autonomous vehicles, Ericsson provides high-capacity RAN for low-latency urban coverage, Nokia ensures reliable transport with energy-efficient networking, and AWS offers a scalable cloud-native core. This combination ensures stringent SLA requirements for real-time communication.
As a result, the optimization process should consider both the decomposition of SLAs and the selection of providers across domains to ensure more flexible and efficient resource utilization in multi-domain networks.
To address these limitations, this paper introduces Risk-Aware Iterated Local Search (RAILS), a novel risk model-driven meta-heuristic framework. RAILS extends the principles of Iterated Local Search (ILS) by integrating dynamic risk modeling and SLA decomposition techniques proposed in~\cite{hsu2024online}.
The main contributions of this paper are:
\begin{enumerate}
    \item [1.] We formulate the joint SLA decomposition and service provider selection tasks as a Mixed-Integer Nonlinear Programming (MINLP) problem and demonstrate its NP-hardness.
    \item [2.] We propose RAILS, a novel risk-aware meta-heuristic framework designed to jointly address the SLA decomposition and service provider selection problem.
    \item [3.] We empirically show that RAILS achieves near-optimal performance within an analytic model-based simulation environment with low computational overhead.
\end{enumerate}
The paper is organized as follows: Section II defines the system model and formulates the problem. Section III introduces the RAILS framework. Section IV details the simulation setup, while Section V presents and discusses the results. Section VI concludes the paper.
%These risk models, continuously updated through real-time feedback mechanisms, guide the local search process to explore provider selections and delay allocations that maximize the E2E acceptance probability. This joint optimization approach ensures that the SLA decomposition not only adheres to performance constraints but also dynamically adjusts to evolving network conditions.

% \section{Related Work}
% \textcolor{red}{\blindtext[1]}
% \textcolor{red}{\blindtext[1]}

\section{System Model}
% \section{Problem Formulation}

% In modern communication networks, the increasing demand for diverse and quality-assured services has motivated the development of network slicing. In a real-world network system, a network slice may span multiple technology domains (e.g., radio access, transport, and core networks), each managed by a set of local service providers. To guarantee the overall service quality, an end-to-end (E2E) Service Level Agreement (SLA) must be decomposed into partial SLAs that are enforced by each domain. This decomposition involves both the selection of an appropriate service provider within each domain and the allocation of a portion of the global delay budget to that domain.

% We consider a two-level network slice management architecture. At the top level, an E2E service orchestrator is responsible for ensuring that the overall SLA is met, whereas at the lower level, local domain controllers manage their respective domains. Due to practical constraints, the orchestrator has access only to historical feedback of admission control from the domain controllers; these data are used to construct risk models that capture each provider's acceptance probability as a function of the delay budget allocated. Given an E2E delay budget \( d_{\text{e2e}} \), the orchestrator must (i) select one service provider per domain and (ii) decompose the E2E delay budget into domain-specific delay SLO such that the E2E acceptance probability is maximized.

\subsection{Problem Formulation}
\label{sec:problem formulation}

Let \( N \) denote the number of domains that the service spans. For each domain \( i \) (with \( i=1,\dots,N \)), let \( \mathcal{J}_i \) denote the set of available service providers. We define the following decision variables:
\begin{itemize}
    \item \( x_{ij} \in \{0,1\} \): a binary variable that is $1$ if provider \( j \in \mathcal{J}_i \) is selected for domain \( i \), and $0$ otherwise.
    \item \( d_i \ge 0 \): the portion of the E2E delay budget allocated to domain \( i \).
\end{itemize}
The acceptance probability of domain \( i \) using provider \( j \) when allocated a delay of \( d_i \) is given by the function \( p_{ij}(d_i) \).
Assuming that the decisions made in the domains are statistically independent, the E2E acceptance probability \( p_{e2e} \) is modeled as the product of the acceptance probabilities of all domains:
\begin{equation}\label{eq:e2eprob}
p_{\text{e2e}} = \prod_{i=1}^{N} \left( \sum_{j \in \mathcal{J}_i} x_{ij}\, p_{ij}(d_i) \right).
\end{equation}
The goal is to \textbf{choose a provider in each domain} and \textbf{allocate the delay budgets} \( \{d_i\}_{i=1}^N \) such that the \textbf{E2E acceptance probability is maximized}, subject to the constraint that the domain-specific partial delays sum up to the E2E delay budget \( d_{\text{e2e}} \). Formally, the problem is formulated as follows:
\begin{equation}\label{eq:optimization_problem}
\begin{aligned}
\max_{\{x_{ij},\,d_i\}} \quad & \prod_{i=1}^{N} \left( \sum_{j \in \mathcal{J}_i} x_{ij}\, p_{ij}(d_i) \right) \\
\text{s.t.} \quad & \sum_{i=1}^{N} d_i = d_{\text{e2e}}, \\
& d_i \ge 0, \quad \forall\, i=1,\dots,N, \\
& \sum_{j \in \mathcal{J}_i} x_{ij} = 1, \quad \forall\, i=1,\dots,N, \\
& x_{ij} \in \{0,1\}, \quad \forall\, i=1,\dots,N,\; \forall\, j \in \mathcal{J}_i.
\end{aligned}
\end{equation}

The presence of both integer and continuous variables, coupled with the nonlinear characteristics of the objective function, designates the problem as a canonical MINLP problem. However, due to the lack of knowledge about \( p_{ij}(d_i) \), we leverage historical feedback data to construct a NN-based risk model for each domain~\cite{SLADNN23}. These risk models serve as surrogates for \( p_{ij}(d_i) \) in the optimization process, providing an estimated acceptance probability based on past observations.

\subsection{NP-Hardness Analysis}
We demonstrate that the joint optimization problem described in~(\ref{eq:optimization_problem}) is NP-hard by reducing from the well-known Multiple-Choice Subset Sum Problem (MCSSP). To this end, we consider a simplified version of the problem where only one provider is available per domain. In this case, the decision problem simplifies to the following objective:
\begin{equation}\label{eq:simple_problem}
\begin{aligned}
\max_{d_i} \quad & \prod_{i=1}^N p_i(d_i)\\
\text{s.t.} \quad & \sum_{i=1}^{N} d_i = d_{\text{e2e}}, \\
& d_i \ge 0, \quad \forall\, i=1,\dots,N.
\end{aligned}
\end{equation}
Now, we define an instance of MCSSP: Given $N$ disjoint sets $\{S_i\}_{i=1}^N$ of nonnegative integers and a target sum $K$, the goal is to determine whether it is possible to choose exactly one element from each set such that the sum of the selected elements equals $K$.
We then construct a corresponding instance of the simplified problem described in~(\ref{eq:simple_problem}) as follows:
\begin{itemize}
    \item The number of domains is set to $N$.
    \item The total E2E delay budget is $d_{\text{e2e}} = K$.
    \item For the provider in each domain $i$, we define the acceptance probability function $p_i(d_i)$ as:
    \begin{equation}
    p_i(d_i) =
    \begin{cases}
        1, & \text{if } d_i \in S_i, \\
        0, & \text{otherwise}.
    \end{cases}
    \end{equation}
    \item The decision version of the simplified problem asks \textit{whether there exists a feasible delay allocation $\{d_i\}_{i=1}^N$ such that $p_{\text{e2e}}=1$, subject to the constraints in}~(\ref{eq:simple_problem}).
\end{itemize}
This construction ensures that $p_i(d_i) = 1$ if and only if the allocated delay $d_i$ is equal to at least one element in $S_i$. As a result, $p_{\text{e2e}}$ is equal to 1 if and only if $d_i \in S_i$ for all $i$. Moreover, since $\{d_i\}_{i=1}^N$ sum up to $K$, any feasible solution that achieves $p_{\text{e2e}} = 1$ corresponds to a valid solution to the MCSSP instance.
%Conversely, if no such selection exists, then at least one domain will be assigned a value $d_i \notin S_i$, resulting in $p_i(d_i) = 0$ and hence $p_{\text{e2e}} = 0$.
Therefore, solving the decision version of the simplified problem is equivalent to solving MCSSP, which is known to be NP-complete~\cite{Kellerer2004}. Hence, the optimization version of simplified problem in~(\ref{eq:simple_problem}) is NP-hard. Since the original problem in~(\ref{eq:optimization_problem}) generalizes this setting by incorporating continuous, nonlinear acceptance probability functions and allowing multiple providers per domain, it is also NP-hard.

Given the NP-hard nature of the problem, finding an exact solution is computationally intractable for large-scale systems. As a result, we resort to meta-heuristic approaches, leveraging the domain-specific risk models built from historical data to guide the search for near-optimal solutions.% In the subsequent section, we describe the proposed meta-heuristic approach.

\section{Methodology}

\subsection{Background}
\noindent \textbf{ILS.} Iterated Local Search (ILS)~\cite{Lourenço2003} is a meta-heuristic approach that enhances local search algorithms by escaping local optima through iterative perturbations and refinements.
% It is especially useful for combinatorial optimization problems, where the search space is large and complex.
ILS operates by first generating an initial solution, either randomly or via a heuristic method, and then refining it through a local search procedure to find a local optimum. Once a local optimum is identified, the algorithm introduces controlled randomness to perturb the solution and push it away from the identified optimum.
% A predefined acceptance criterion is then applied to determine whether the perturbed solution should replace the current solution.
This cycle of local search and perturbation continues until a stopping condition, such as reaching a maximum number of iterations or meeting a convergence criterion, is satisfied.
% \noindent \textbf{ILS.} Iterated Local Search (ILS)~\cite{Lourenço2003} is a meta-heuristic approach that enhances local search algorithms by escaping local optima through iterative perturbations and refinements. ILS is particularly useful for combinatorial optimization problems, where the search space is large and complex. ILS operates by iteratively refining solutions through local search and perturbation steps. The key components of the algorithm are:
% \begin{itemize}
%     \item \textbf{Initial Solution.} A starting solution is generated, either randomly or through a heuristic approach.
%     \item \textbf{Local Search.} The solution is refined using a local search technique to find a local optimum.
%     \item \textbf{Perturbation.} The local optimum is modified by introducing controlled randomness, pushing the solution away from the local optima.
%     \item \textbf{Acceptance Criterion.} The perturbed solution is accepted based on predefined criteria, such as improvements or probabilistic acceptance.
%     \item \textbf{Iteration.} The process repeats until a stopping criterion, such as a maximum number of iterations or convergence is met.
% \end{itemize}
ILS is used in networked cloud resource mapping to address the challenge of optimally partitioning and embedding virtual resources across multiple cloud providers~\cite{6226390}. ILS-based request partitioning has been shown to effectively balance cost and performance, leading to improved virtual network embedding outcomes.

\noindent \textbf{RADE.} Real-time Adaptive DEcomposition (RADE)~\cite{hsu2024online} is an advanced SLA decomposition framework that dynamically adjusts decomposition strategies based on real-time feedback of network conditions. Unlike static decomposition approaches, RADE employs online learning to enhance adaptability and accuracy. It utilizes a two-step decomposition approach~\cite{Vleeschauwer21_SLAdecomposition, SLADNN23}. First, the orchestrator maintains domain-specific NN-based risk models trained on historical SLA acceptance and rejection feedback. Next, the E2E SLA is decomposed into domain-specific SLAs to maximize the overall acceptance probability, using a grid search followed by Sequential Least Squares Programming (SLSQP) algorithm. To adapt to evolving network conditions, these risk models are updated timely via Online Gradient Descent (OGD). A First In First Out (FIFO) memory buffer preserves recent observations, ensuring stable learning while mitigating overfitting caused by transient anomalies.
RADE addresses key limitations of static decomposition methods by incorporating real-time adaptation. It also offers resilience against data corruption through its FIFO memory buffer. Experimental results show that RADE consistently outperforms traditional methods in dynamic multi-domain environments, making it a promising solution for adaptive SLA management in modern network architectures.
% \noindent \textbf{RADE.}
% Real-time Adaptive DEcomposition (RADE)~\cite{hsu2024online} is an advanced SLA decomposition framework that dynamically adjusts decomposition strategies based on real-time feedback and network conditions. Unlike static decomposition approaches, RADE incorporates online learning to enhance adaptability and accuracy.
% RADE employs a two-step decomposition approach combined with neural network-based risk models:

% \begin{itemize}
%     \item \textbf{Risk Model Learning.} The orchestrator maintains domain-specific risk models trained on historical SLA acceptance and rejection feedback.
%     \item \textbf{Optimization-based Decomposition.} The E2E SLA is decomposed into domain-specific SLAs to maximize the E2E acceptance probability. A grid search followed by the SLSQP algorithm is employed.
%     \item \textbf{Online Update.} The risk models are updated in real-time using Online Gradient Descent (OGD) to adapt to evolving network conditions.
%     \item \textbf{FIFO Memory Buffer.} A memory buffer maintains recent observations, ensuring stable  learning while preventing overfitting from transient anomalies.
% \end{itemize}

% RADE addresses key limitations of static decomposition methods by incorporating real-time adaptation mechanisms.
% Furthermore, RADE offers resilience against data corruption with FIFO memory buffer.
% Experimental results demonstrate that RADE consistently outperforms traditional methods in dynamic multi-domain environments, making it a promising approach for adaptive SLA management in modern network architectures.

\subsection{Risk-Aware Iterated Local Search}
In this work, we propose RAILS, a risk model-driven meta-heuristic method to solve the joint provider selection and SLA decomposition problem.
The optimization problem involves two interconnected sets of decision variables (see Section~\ref{sec:problem formulation}). On one hand, we have discrete variables that determine which provider is selected in each domain. On the other hand, we have continuous variables that specify how the E2E delay budget is decomposed among the domains to maximize the E2E acceptance probability. These two aspects of the problem are inherently intertwined because the acceptance probability in each domain is computed using risk models that depend on both the selected provider and the assigned delay requests.
Specifically, once provider $j$ is chosen for domain $i$, the corresponding risk model serves as its surrogate, predicting \( p_{ij}(d_i) \) for a given delay budget \( d_i \).
%The risk models, which are constructed from historical data using NNs, capture the admission control behavior of each provider as a function of the delay allocation.
%When a provider is selected for a domain, the corresponding risk model serves as its surrogate, predicting performance for a given delay budget.
%This means that the evaluation of a potential solution—i.e., a combination of provider selections and delay assignments—cannot be decoupled into two independent problems.
%A provider might appear attractive under a particular delay allocation in one domain, but the overall E2E performance also depends on the delay allocations and provider choices in other domains.
%In other words, the discrete and continuous decisions interact nonlinearly through the acceptance probability functions.
In RAILS, the framework efficiently explores the complex search space by iteratively refining provider selections with risk models.
%Local search operators adjust selections, while up-to-date risk models compute maximum E2E acceptance probabilities, guiding the search.
%The RAILS approach leverages these refinements and risk assessments to navigate the non-separable optimization landscape.

% The ILS framework serves as a robust heuristic for tackling the complex search space generated by these interconnected decision variables. Within each iteration of ILS, local search operators explore the neighborhood of the current solution by adjusting the provider selections. The risk models play a critical role during the evaluation phase of ILS. For each candidate provider selections, the up-to-date risk models are used to compute the corresponding maximum E2E acceptance probabilities using RADE, which in turn guide the search toward more promising regions of the solution space. Thus, the proposed RAILS approach leverages the iterative refinement of candidate solutions, guided by risk model assessments, to navigate the challenging, non-separable optimization landscape.
\begin{figure}[ht]
     \centering
     \includegraphics[width=0.93\columnwidth]{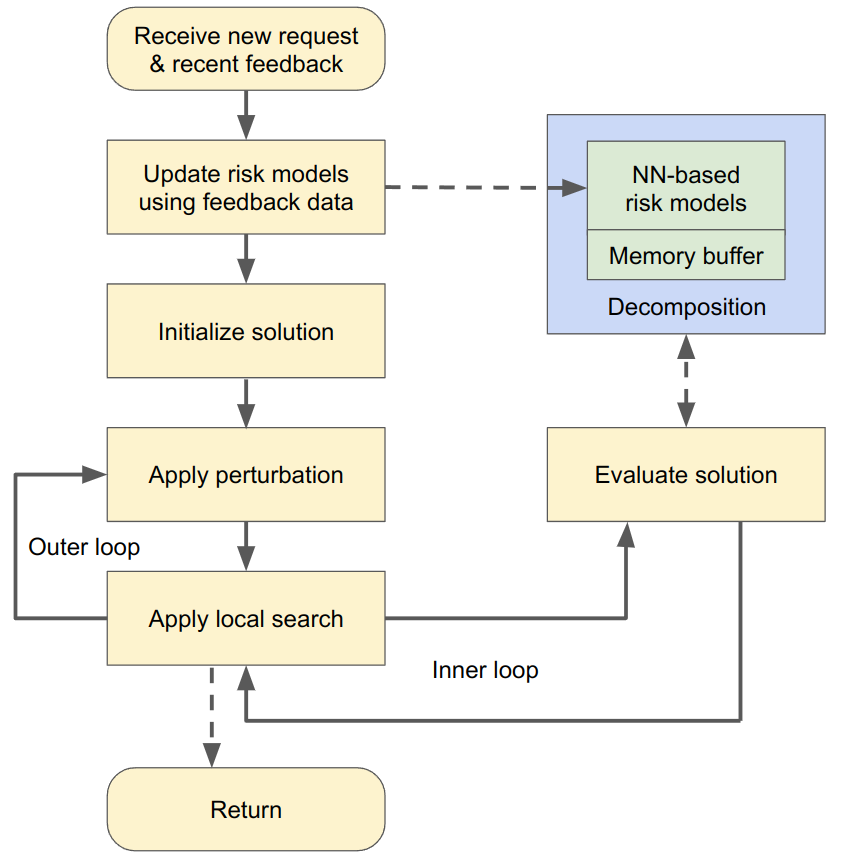}
    \caption{A single-iteration workflow of RAILS.}
    \label{fig:flow}
\end{figure}
Fig.~\ref{fig:flow} illustrates the workflow of the RAILS algorithm. The process begins with updating the risk models and memory buffer using the latest feedback data.
Next, an initial solution for provider selection is generated. The algorithm then enters an iterative refinement phase with perturbation and local search steps.
%These steps leverage the risk models to identify neighboring solutions with the highest E2E acceptance probability.
The perturbation step randomly alters the provider selection for each domain with a probability $p_{\mu}$ to possibly escape local optima, while the local search step refines the solution by randomly selecting a domain and exhaustively checking all provider options within the domain to identify the best one based on the risk models.
This iterative process continues within an inner and outer loop structure until a predefined stopping condition is met, at which point the best solution identified is returned. RAILS operates within an ILS framework, where the core evaluation mechanism is powered by RADE. Specifically, ILS performs the repeated perturbation and local search steps, while RADE handles dynamic risk modeling and real-time decomposition. This synergy enables effective handling of the coupled discrete–continuous nature of provider selection and SLA decomposition.

\section{Performance Evaluation}

\subsection{Simulation Environment}
In our simulation environment, we model the dynamic behavior of each provider's system load and the resulting performance characteristics that affect SLA acceptance. In particular, we capture the temporal variations in load and their impact on the minimum delay that a provider can support, which in turn governs the acceptance probability of an SLA request. Because this subsection focuses on a single-domain provider, we omit the $i$ and $j$ subscripts for clarity.

\noindent \textbf{System Load Modeling.} For each provider, the system load is assumed to evolve periodically over time. Let \( t \) denote the current time. The system load \( \ell(t) \) is modeled using a sinusoidal function~\cite{10.1145/2188286.2188301} as follows:
\begin{equation}\label{eq:system_load}
\ell(t) = \ell_{\text{base}} \cdot k + \ell_{\text{base}} \cdot (1 - k) \cdot \frac{1 + \sin\!\left(\frac{2\pi t}{T} + \phi\right)}{2},
\end{equation}
where \(\ell_{\text{base}}\) is a constant representing the baseline load of the provider, \( k \in [0,1] \) is a parameter that determines the fraction of the load that is static, \( T \) is the period of the sinusoidal fluctuation, and \(\phi\) is the phase shift.
% where:
% \begin{itemize}
%     \item \(\ell_{\text{base}}\) is a constant representing the baseline load of the provider.
%     \item \( k \in [0,1] \) is a parameter that determines the fraction of the load that is static.
%     \item \( T \) is the period of the sinusoidal fluctuation.
%     \item \(\phi\) is the phase shift.
% \end{itemize}
This formulation ensures that the system load varies between the minimum load \(\ell_{\text{base}} \cdot k\)  and the maximum load \(\ell_{\text{base}}\). The parameter $k$ allows for a mixture of a constant baseline load and a dynamic component.
%thereby enabling the simulation of various operational conditions in real-world network systems.

\begin{figure}[ht]
     \centering
     \includegraphics[width=0.9\columnwidth]{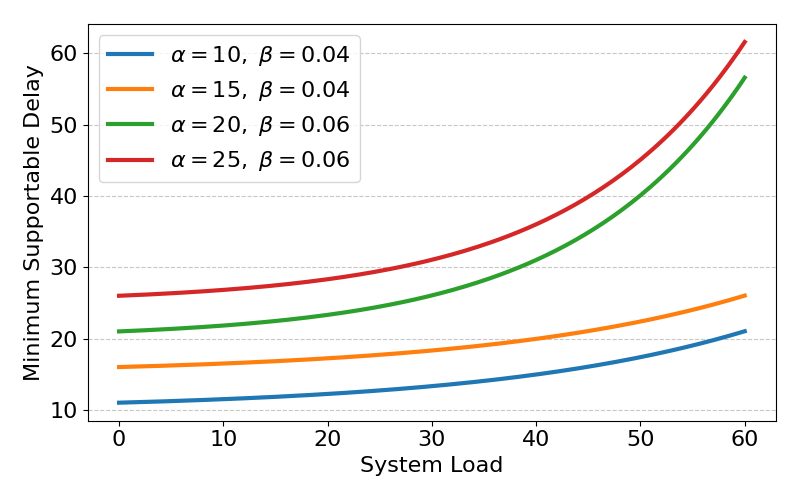}
    \caption{Minimum supportable delay versus system load.}
    \label{fig:delay}
    \vspace{-0.5em}
\end{figure}

\noindent \textbf{Minimum Supportable Delay.}
Given the dynamic system load defined in~(\ref{eq:system_load}), the minimum delay that a provider can support for an incoming request is assumed to depend on both a fixed latency component and an exponential function of the system load. Specifically, the minimum supportable delay \( d^{\min}(t) \) is defined as:
\begin{equation}\label{eq:min_delay}
d^{\min}(t) = \alpha + \exp\!\left(\beta \cdot \ell(t)\right),
\end{equation}
where \( \alpha \) represents the inherent latency of the system when the load is minimal (i.e., the baseline latency), and \(\beta\) is a parameter that characterizes how sensitive the delay is to changes against system load.
% where:
% \begin{itemize}
%     \item \(\alpha\) represents the inherent latency of the system when the load is minimal (i.e., the baseline latency).
%     \item \(\beta\) is a parameter that characterizes how sensitive the delay is to changes against system load.
% \end{itemize}
This expression reflects that as the system load increases, the provider's capability to handle requests with low delay diminishes, leading to a higher minimum supportable delay. Fig.~\ref{fig:delay} demonstrates this effect for different parameter values.
The exponential relationship captures the non-linear increase in minimum supportable delay as the load intensifies, while \(\alpha\) sets the lower bound of latency.

\begin{figure}[ht]
     \centering
     \includegraphics[width=0.9\columnwidth]{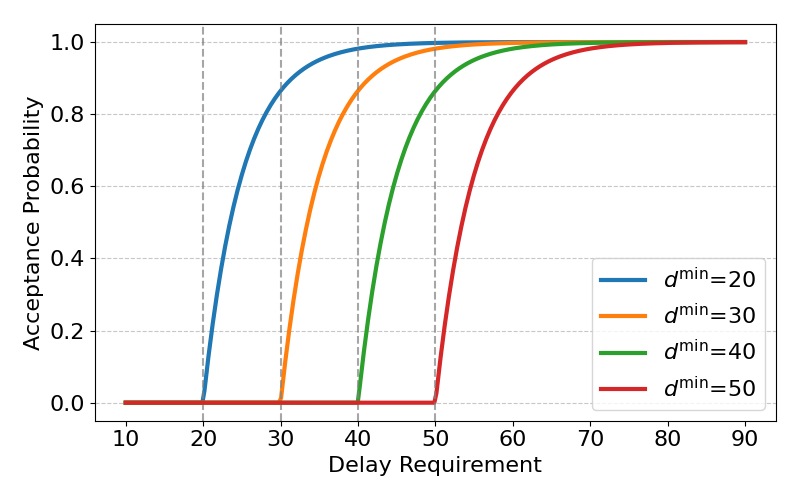}
    \caption{Piecewise acceptance response curves.}
    \label{fig:prob}
    \vspace{-0.5em}
\end{figure}

\noindent \textbf{Acceptance Probability.}
Based on the minimum supportable delay defined in~(\ref{eq:min_delay}), the acceptance probability of a service level request is defined as a function of the requested delay \( d \). Specifically, if the requested delay is less than the minimum supportable delay \( d^{\min}(t) \), the request is rejected. Otherwise, the acceptance probability follows an S-curve, increasing with the excess delay and saturating as the requested delay becomes much larger than \( d^{\min}(t) \)~\cite{Vleeschauwer21_SLAdecomposition}. Formally, the acceptance probability \( p(d; t) \) is defined as:
\begin{equation}\label{eq:acceptance_probability}
p(d; t) =
\begin{cases}
0, & \text{if } d < d^{\min}(t), \\[1mm]
1 - \exp\!\left(-\lambda \, (d - d^{\min}(t))\right), & \text{if } d \ge d^{\min}(t),
\end{cases}
\end{equation}
where \( \lambda > 0 \) is a parameter that controls the rate at which the acceptance probability increases as the delay requirement exceeds the minimum supportable delay. The response curves with different parameter settings are illustrated in Fig.~\ref{fig:prob}.

The piecewise function ensures that any request with a delay requirement less than the system's minimum supportable delay is rejected. For requests above this threshold, the probability of acceptance increases rapidly at first and gradually levels off, reflecting the typical behavior of admission control in real-world service provider systems.
%This simulation framework enables us to evaluate our SLA decomposition strategies under realistic and dynamically varying conditions, providing insights into the performance of our proposed method.

\subsection{Evaluation Scenarios}

To assess the long-term performance of the proposed RAILS framework, we conduct simulations over $100$ discrete time steps within the designed simulation environment. At each time step, a new service request arrives with an E2E delay budget of $100$\,ms. The RAILS is then applied to select a service provider for each domain and to determine the corresponding delay decomposition that maximizes the E2E acceptance probability according to the risk models built from historical data. Specifically, once the RAILS provides a provider selection and delay decomposition, this assignment is evaluated using the ground-truth acceptance probability models (described in~(\ref{eq:acceptance_probability})) to compute the actual E2E acceptance probability. This process is repeated at every time step, and the resulting E2E acceptance probabilities are collected to compute an overall average performance metric, namely:
\begin{equation}\label{eq:avge2e}
\overline{p}_{\text{e2e}} = \frac{1}{T} \sum_{t=1}^{T} p_{\text{e2e}}^{(t)} = \frac{1}{T} \sum_{t=1}^{T} \prod_{i=1}^{N} \left( \sum_{j \in \mathcal{J}_i} x_{ij}^{(t)} \, p_{ij}^{(t)}(d_i^{(t)}) \right),
\end{equation}
where \( T \) denote the total number of simulation time steps, and \( N \) represents the total number of involved domains.

% Let \( T \) denote the total number of simulation time steps (e.g., \(T=100\)). At each time step \( t \) (\( t = 1, 2, \ldots, T \)), suppose that an assignment is made by selecting one provider per domain and decomposing the E2E delay budget among the \( N \) domains. For domain \( i \) at time \( t \), let:
% \begin{itemize}
%     \item \( x_{ij}^{(t)} \in \{0,1\} \) be the binary variable that is 1 if provider \( j \) (from the set \( \mathcal{J}_i \)) is selected for domain \( i \) at time \( t \), and 0 otherwise,
%     \item \( d_{t,i} \ge 0 \) be the delay budget allocated to domain \( i \) at time \( t \),
%     \item \( p_{ij}(d_{t,i}) \) be the acceptance probability of domain \( i \) when using provider \( j \) given a delay \( d_{t,i} \).
% \end{itemize}

% Thus, the effective acceptance probability for domain \( i \) at time \( t \) is given by:
% \[
% P_{i}^{(t)} = \sum_{j \in \mathcal{J}_i} x_{ij}^{(t)} \, p_{ij}(d_{t,i}).
% \]

% Assuming that the acceptance events in the domains are independent, the overall E2E acceptance probability at time \( t \) is the product of the individual domain acceptance probabilities:
% \[
% P_{\text{E2E}}^{(t)} = \prod_{i=1}^{N} P_{i}^{(t)} = \prod_{i=1}^{N} \left( \sum_{j \in \mathcal{J}_i} x_{ij}^{(t)} \, p_{ij}(d_{t,i}) \right).
% \]

% Finally, the overall E2E average acceptance probability over the \( T \) time steps is defined as:
% \[
% \overline{P}_{\text{E2E}} = \frac{1}{T} \sum_{t=1}^{T} P_{\text{E2E}}^{(t)} = \frac{1}{T} \sum_{t=1}^{T} \prod_{i=1}^{N} \left( \sum_{j \in \mathcal{J}_i} x_{ij}^{(t)} \, p_{ij}(d_{t,i}) \right).
% \]

At each time step, we assume that a set of historical requests and their associated feedback are available from each provider. A historical sample is represented by a pair \((d^{(t)}, a^{(t)})\), where \(d^{(t)}\) is the delay request and \(a^{(t)} \in \{0,1\}\) is the binary decision outcome from the admission control. For generating the feedback data, we simulate requests by sampling delay requirements uniformly from the interval \([10\,\text{ms}, 100\,\text{ms}]\). Each request is then processed through the corresponding ground-truth acceptance probability model to obtain its actual acceptance probability, and a coin-flipping process is used to determine whether the request is accepted or not by the admission control.
For performance comparison, we consider two baselines:\\
% \begin{itemize}
%     \item \textbf{Non-Risk-Aware (NRA).} In this baseline, a provider is selected at random for each domain, and the E2E delay budget is evenly decomposed among the domains as a heuristic guess, without leveraging risk models.
%     \item \textbf{Optimal (OPT).} This benchmark is obtained via an exhaustive search with full access to ground-truth acceptance probability models, representing the theoretical upper bound on performance.
% \end{itemize}
\noindent \textbf{Non-Risk-Aware (NRA).} In this baseline, a provider is selected at random for each domain, and the E2E delay budget is evenly decomposed among the domains as a heuristic guess, without leveraging risk models.\\
\noindent \textbf{Optimal (OPT).} This benchmark uses an exhaustive search with full access to ground-truth acceptance probability models, representing the theoretical upper bound on performance.

%By comparing the RAILS approach with these baselines, we aim to demonstrate its effectiveness in  achieving higher E2E acceptance probabilities over time, while maintaining reasonable computational time.

\subsection{Experimental Setup}

In our experiments, we consider a network slicing scenario involving $3$ domains, each comprising 10 service providers. The ground-truth models for each provider are generated using a set of randomly selected parameters to emulate realistic and heterogeneous operational conditions. For each provider, the baseline latency \(\alpha\) is drawn uniformly from \([0, 2]\); The parameter $\lambda$ in~(\ref{eq:acceptance_probability}) is set to $0.2$, and $k$ is set to $0.5$ in~(\ref{eq:system_load}); A domain-wise additional latency, randomly chosen from the set \(\{0, 10, 20\}\), is added to the baseline latency \(\alpha\) to reflect inter-domain behavioral shifts; the load-sensitivity parameter \(\beta\) is sampled uniformly from \([0.04, 0.06]\); the baseline system load \(\ell_{\text{base}}\) is drawn uniformly from \([30, 50]\); the period of the sinusoidal load fluctuation is selected as an integer uniformly from the range \([30, 60]\); and the phase shift in the load function is chosen uniformly from the interval \([0, \pi]\).

At each time step, we assume that the number of recent feedback samples available from each provider is proportional to its current system load defined in~(\ref{eq:system_load}).  We use the integer part of the load as the number of samples. For the RAILS framework, the number of iterations is set empirically to the total number of providers across all domains (i.e., \(3 \times 10 = 30\)), and the perturbation probability $p_{\mu}$ is $0.8$. Results are averaged over $10$ independent runs to ensure statistical significance.
Each risk model is implemented as a $2$-layer monotonic Multi-Layer Perceptron (MLP) with a hidden dimension of $16$,
%Layer Normalization (LN), and ReLU activation function,
similar to that described in~\cite{hsu2024online}. AdamW optimizer is used with a learning rate of $0.01$. A memory buffer of size $300$ is maintained, and each risk model update involves $10$ iterations.

\section{Results and Discussion}
\begin{figure}[h]
    \centering
    \begin{subfigure}{0.3\textwidth}
        \centering
        \includegraphics[width=\textwidth]{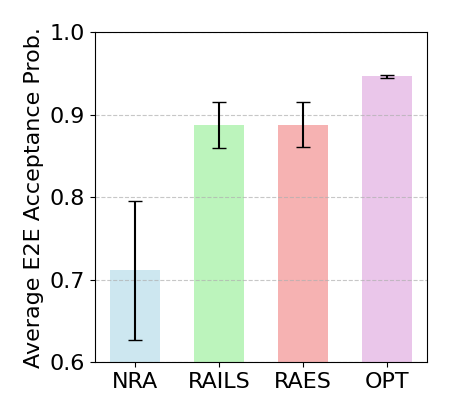} % Replace with your first image file path
        % \caption{Run Time Comparison between RAILS and RAES}
        \caption{E2E Acceptance Probability}
        \label{fig:e2eacc}
    \end{subfigure}
    \begin{subfigure}{0.175\textwidth}
        \centering
        \includegraphics[width=\textwidth]{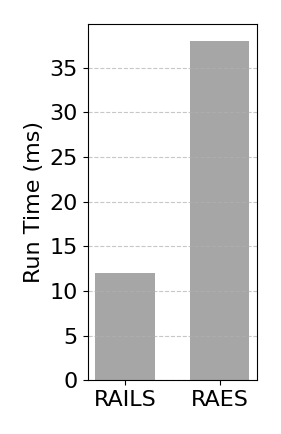} % Replace with your second image file path
        % \caption{E2E Acceptance Probability for NRA, RAILS, RAES, and OPT}
        \caption{Run Time}
        \label{fig:runtime}
    \end{subfigure}
    \caption{Performance comparison: (a) E2E SLA acceptance probability over time; (b) computational run time.}
    \label{fig:combined}
\end{figure}

% Average E2E acceptance probability over time
Fig.~\ref{fig:e2eacc} presents the average E2E acceptance probability for the considered approaches. Each bar reflects the average performance metric defined in~(\ref{eq:avge2e}).
%, while Fig.~\ref{fig:runtime} reports the average run time per request. 
The NRA approach achieves an average acceptance probability of approximately $0.71$. The large error bars reveal significant performance fluctuations, suggesting that the absence of strategic decision-making leads to suboptimal performance. In contrast, the proposed RAILS method demonstrates a substantial improvement, achieving an average acceptance probability of around $0.89$. The error bars are smaller compared to the NRA approach, indicating more stable outcomes.
Moreover, we include results for a RAILS variant, denoted RAES, where the ILS search component is replaced with an exhaustive search. RAES's performance reflects RAILS running with an effectively infinite number of iterations. Both RAILS and RAES achieve comparable performance, while RAILS requires only $31.6\%$ of RAES's run time, demonstrating its computational efficiency, as shown in Fig.~\ref{fig:runtime}.
The OPT approach, which represents the theoretical performance upper bound, achieves an average acceptance probability of about $0.95$. While RAILS does not fully reach the OPT benchmark, it comes remarkably close, balancing computational efficiency with SLA acceptance rates.
% \begin{figure}[h]
%      \centering
%      \includegraphics[width=0.8\columnwidth]{figures/e2eacc.png}
%     \caption{Average E2E acceptance probabilities over time (colored bars, left axis) and run time (gray bars, right axis).}
%     \label{fig:e2eacc}
%     % \vspace{-0.5em}
% \end{figure}

% \begin{table}
% \centering
% \caption{Partial trace of a simulation run.}\label{tab:log}
% \begin{tabular}{|c|c|c|c|c|}
% \hline
% Step & Method & $p_{\text{e2e}}$ & $d_1, d_2, d_3$ & Providers \\
% \hline
% \multirow{3}{*}{1} & NRA & 0.47 & 33.33, 33.33, 33.33 & 2, 4, 5 \\
%  & RAILS & 0.93 & 25.93, 35.02, 40.07 & 1, 0, 3 \\
%  & OPT & 0.94 & 24.59, 33.63, 42.78 & 9, 1, 3 \\
% \hline
% \multirow{3}{*}{5} & NRA & 0.71 & 33.33, 33.33, 33.33 & 3, 3, 6 \\
%  & RAILS & 0.93 & 23.89, 33.56, 43.55 & 7, 0, 8 \\
%  & OPT & 0.95 & 23.78, 33.55, 42.67 & 9, 6, 3 \\
% \hline
% \multirow{3}{*}{10} & NRA & 0.68 & 33.33, 33.33, 33.33 & 8, 8, 9 \\
%  & RAILS & 0.94 & 23.78, 32.20, 44.02 & 0, 0, 8 \\
%  & OPT & 0.95 & 23.55, 33.21, 43.24 & 9, 6, 3 \\
% \hline
% \multirow{3}{*}{15} & NRA & 0.84 & 33.33, 33.33, 33.33 & 2, 4, 8 \\
%  & RAILS & 0.93 & 22.91, 34.06, 43.03 & 0, 3, 8 \\
%  & OPT & 0.95 & 23.66, 33.22, 43.13 & 9, 3, 3 \\
% \hline
% \multirow{3}{*}{20} & NRA & 0.79 & 33.33, 33.33, 33.33 & 7, 5, 8 \\
%  & RAILS & 0.94 & 19.99, 36.75, 43.26 & 2, 7, 8 \\
%  & OPT & 0.95 & 23.82, 33.11, 43.08 & 9, 7, 1 \\
% \hline
% \end{tabular}
% \end{table}
% trace of selected providers
\begin{table}[ht]
\centering
\caption{Partial trace of a simulation run.}\label{tab:log}
\begin{tabular}{|c|c|c|c|c|}
\hline
Step & Method & $p_{\text{e2e}}$ & $d_1, d_2, d_3$ & Providers \\
\hline
\multirow{3}{*}{0} & NRA & 0.78 & 33.33, 33.33, 33.33 & 7, 9, 6 \\
& RAILS & 0.92 & 22.00, 33.00, 45.00 & 8, 6, 0 \\
& OPT & 0.94 & 24.23, 32.50, 43.27 & 3, 1, 0 \\
\hline
\multirow{3}{*}{5} & NRA & 0.81 & 33.33, 33.33, 33.33 & 3, 4, 9 \\
& RAILS & 0.88 & 22.63, 32.08, 45.30 & 4, 6, 1 \\
& OPT & 0.94 & 24.31, 32.49, 43.20 & 3, 1, 6 \\
\hline
\multirow{3}{*}{10} & NRA & 0.75 & 33.33, 33.33, 33.33 & 5, 4, 1 \\
& RAILS & 0.93 & 26.22, 30.30, 43.48 & 8, 6, 6 \\
& OPT & 0.94 & 24.17, 32.98, 42.85 & 5, 1, 3 \\
\hline
\multirow{3}{*}{15} & NRA & 0.73 & 33.33, 33.33, 33.33 & 7, 3, 9 \\
& RAILS & 0.90 & 20.00, 30.00, 50.00 & 1, 6, 7 \\
& OPT & 0.94 & 24.11, 33.00, 42.90 & 2, 4, 3 \\
\hline
\end{tabular}
\end{table}

Table~\ref{tab:log} provides a partial trace of a simulation run, showcasing the E2E acceptance probabilities, delay decompositions, and the indices of the selected providers across different time steps. The OPT method shows that the optimal provider selections and delay decompositions change frequently over time, reflecting dynamic network conditions. For instance, optimal providers shift from $(3, 1, 0)$ at step $0$ to $(2, 4, 3)$ at step $15$, highlighting the need for adaptive risk models to maintain high SLA acceptance rates.
Furthermore, RAILS achieves near-optimal performance across time steps, even without always selecting the optimal providers.
%Even without always selecting the optimal providers, RAILS achieves E2E acceptance probabilities close to the Optimal method across time steps.
\begin{figure}[ht]
     \centering
     \includegraphics[width=0.9\columnwidth]{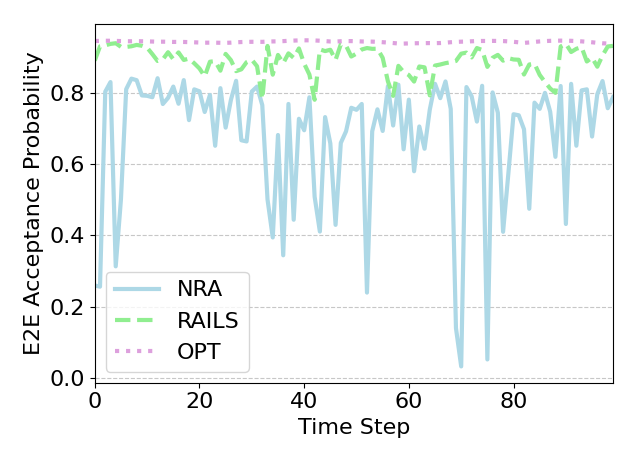}
    \caption{E2E acceptance probability over time for a single run.}
    \label{fig:e2eacc_run}
    % \vspace{-0.5em}
\end{figure}

Fig.~\ref{fig:e2eacc_run} illustrates the temporal dynamics of E2E acceptance probabilities for a single representative run. The OPT method consistently achieves near-perfect acceptance.
RAILS closely tracks the optimal performance, with slight fluctuations, indicating its effectiveness in adapting to dynamic network conditions.
In contrast, the NRA method experiences significant variability and frequent drops in acceptance probability, revealing its limitations in predicting and responding to environmental changes.
% complexity analysis?

\section{Conclusion}
This paper introduced Risk-Aware Iterated Local Search (RAILS), a meta-heuristic framework driven by NN-based risk models, for SLA decomposition and service provider selection in multi-domain networks. We formulated the problem as a Mixed-Integer Nonlinear Programming (MINLP) problem and demonstrated its NP-hardness. RAILS integrates dynamic risk modeling with iterated local search, effectively handling the complex optimization landscape of interdependent decisions. Simulation results showed that RAILS achieves near-optimal performance against the theoretical optima while maintaining low computational overhead.
%This highlights RAILS' ability to adapt to dynamic network conditions with high SLA acceptance rates efficiently.
Overall, RAILS offers a robust and efficient solution for adaptive network slicing management in modern network systems. Future work will explore the long-term impact of each decision on subsequent ones by formulating the problem as a Markov Decision Process (MDP) and applying Deep Reinforcement Learning (DRL) techniques.
\section*{Acknowledgment}
This research was partially funded by the HORIZON SNS JU DESIRE6G project (grant no. 101096466) and the Dutch 6G flagship project ``Future Network Services''.


\bibliographystyle{IEEEtran}
\bibliography{references.bib}

%\bibliographystyle{IEEEtran}


% \end{thebibliography}

\end{document}